\documentclass[%
 reprint,
 superscriptaddress,
 amsmath,amssymb,
 aps,
 prd,
]{revtex4-1}

\usepackage{dcolumn}
\usepackage{epsfig}
\usepackage{subfigure}
\usepackage{bm}
\usepackage{breakurl}
\usepackage{enumitem}
\usepackage{mdframed}

\newcommand\fig[1]     {Fig.\,{\ref{#1}}}
\usepackage{color}

\newcommand{\Fbeta}{\mbox{\boldmath $\beta$}}

\newcommand{\beq}{\begin{equation}}
\newcommand{\eeq}{\end{equation}}
\newcommand{\bea}{\begin{eqnarray}}
\newcommand{\eea}{\end{eqnarray}}

\newcommand{\nn}{\nonumber\\}

\def\eq#1{(\ref{#1})}
\def\s0#1#2{\mbox{\small{$ \frac{#1}{#2} $}}}
\def\0#1#2{\frac{#1}{#2}}

\begin{document}
\title{Thermal RG Flow of AS Quantum Gravity}

\author{E.~Nyergesy}
\affiliation{University of Debrecen, Institute of Physics, P.O.Box 105, H-4010 Debrecen, Hungary}

\author{I.~G.~M\'ari\'an}
\affiliation{University of Debrecen, Institute of Physics, P.O.Box 105, H-4010 Debrecen, Hungary}
\affiliation{HUN-REN Atomki, P.O.Box 51, H-4001 Debrecen, Hungary} 

\author{E.~Meskhi}
\affiliation{University of Bonn, Department of Physics and Astronomy, Nusallee 12, 53115 Bonn, Germany}

\author{Y.~Turovtsi-Shiutev}
\affiliation{Uzhhorod National University, 14, Universytetska str., Uzhhorod, 88000, Ukraine}

\author{I.~N\'andori}
\affiliation{University of Debrecen, Institute of Physics, P.O.Box 105, H-4010 Debrecen, Hungary}
\affiliation{University of Miskolc, Institute of Physics and Electrical Engineering, H-3515, Miskolc, Hungary}
\affiliation{HUN-REN Atomki, P.O.Box 51, H-4001 Debrecen, Hungary}

\date{\today}

\begin{abstract}
We perform the thermal Renormalization Group (RG) study of the Asymptotically Safe (AS) quantum 
gravity in the Einstein-Hilbert truncation by relating the temperature parameter to the running RG scale 
as $T \equiv k_T = \tau k$ (in natural units) in order to determine its thermal evolution in terms of 
the dimensionless temperature $\tau$ which is associated with the temperature of the expanding 
Universe. Thus, $k_T$ and $k$ are understood as running cutoffs for thermal and quantum 
fluctuations, respectively. Quantum effects are taken into account by moving along the thermal 
RG trajectory with fixed value of $\tau$ producing the quantum effective action at a given 
dimensionless temperature. The $\tau$-evolution of the dimensionless Newton coupling $g(\tau)$ 
and the dimensionless cosmological constant $\lambda(\tau)$ results in a vanishing $g$-coordinate 
of the Reuter (i.e., non-Gaussian UV) fixed point in the high temperature limit ($\tau \to \infty$) which 
means that only the symmetric phase of AS gravity survives at $\tau = \infty$.
Thus, in case of large temperatures the cosmological constant takes on a negative value 
in the limit $k\to 0$ which was also initially predicted by certain string theories, however, in our approach 
this is not in disagreement with observations, since during the thermal evolution of the Universe a phase 
transition occurs and the cosmological constant runs to the expected positive value at low temperatures.
\end{abstract}


\maketitle

\section{Introduction} 
The main point of this work is to apply a modified thermal Renormalization Group (RG) approach 
recently introduced in \cite{thermal_rg_cosmo} on the Asymptotically Safe (AS) quantum gravity, 
see e.g., \cite{reuter,AS_grav_1,lauscher_2002,AS_grav_2,AS_grav_3,AS_grav_4,AS_grav_5,AS_grav_6,
AS_grav_7,AS_grav_8,AS_grav_9} by using the Einstein Hilbert truncation in $d=4$ dimensions. 
Since RG method, in particular the Functional RG (FRG) approach \cite{eea_rg} has been used in 
cosmology very frequently \cite{cosmo_1,cosmo_2,cosmo_3,cosmo_4,cosmo_5,cosmo_6,cosmo_7}, 
the modified thermal RG approach \cite{thermal_rg_cosmo} could naturally find application in 
quantum cosmology, too. It solves an inherent riddle in the application of RG and finite temperature 
quantum field theory techniques. From one side one knows i) that one has to extend cosmological 
models to finite temperatures in which different energy and time scales are connected by the RG flow 
-- from the other side ii) one of the first extension of the FRG method to finite temperatures 
\cite{rg_dimful_T_1} and also all further thermal FRG investigations based on this assumption, 
have problems in determining non-trivial fixed points of the RG flow which are crucial to consider 
phase transitions.

Rationale of the point i) is that although particle physics and cosmology are intimately connected, 
in accelerators a limited number of colliding particles are taken into account, but the early Universe 
must be seen as a hot dense plasma. Thus, the zero temperature quantum field theory which 
is the usual theoretical framework for particle processes must be extended to finite temperatures 
for cosmological applications.

Regarding point ii), thermal field theory is a very well-developed framework for finite temperature 
applications \cite{kapusta,negele,sachdev}. However, there are open question on its connection to 
renormalization and the RG method. By using a natural unit in which $c = \hbar = k_B = 1$, the 
temperature parameter has to be related to a momentum scale. In the perturbative RG approach 
the temperature $T$ is set to be equal to the running RG momentum scale $\mu$, i.e., $\mu = 2\pi T$
while in non-perturbative approach, it is linked to a fixed momentum $\Lambda$, i.e., $T = \tau \Lambda$ 
where $\tau$ is a variable dimensionless parameter, see i.e., one of the very first publication on 
thermal FRG  \cite{rg_dimful_T_1} which served as a reference for thermal FRG applications.
The latter choice is motivated by the fact that the RG scale-dependence is introduced artificially in 
the Wilsonian approach \cite{wilson} and the quantized theory must be obtained in the physical 
limit where the RG scale is sent to zero, $k\to 0$, thus it seems reasonable to connect the 
temperature to a fixed momentum. Indeed, this type of non-perturbative thermal FRG equation is 
well-known and used in a huge number of works with various applications. Some other thermal FRG 
approaches, see e.g., \cite{rg_dimless_T_7,rg_dimless_T_9,rancon_3,rg_dimless_T_11},
were using a dimensionless temperature, $\tau_k = T/k$ which was introduced as a technical tool with
the important constraint: the dimensionful temperature parameter $T$ is kept constant over the RG flow,
so, $\tau_k$ has a trivial scale dependence. For example in \cite{rg_dimless_T_9} at page 42, 
in Eq.(132) one finds $\partial_t \tau = - \tau$ where $t = \ln(k/\Lambda)$ and in \cite{rg_dimless_T_11} 
at the top of the page 12 one can read $\partial_t T/k = - T/k$ which result in the trivial RG scale 
dependence for $\tau_k$. Another example is found in \cite{rg_dimless_T_7} at page 4 above 
equation (21), $\tilde T(\Lambda^\star) = 1$ where the intermediate momentum scale $\Lambda^\star$ 
is fixed by the running dimensionless temperature $\tilde T$ which is identical to $\tau$. Finally, we 
refer to \cite{rancon_3} at page 7 right below equation (50), where one finds $\tilde T_k = T/(c_k k)$. 
Thus, even in these works the temperature parameter was practically linked to a fixed UV momentum 
cutoff $\Lambda$, i.e., $T = \tau \Lambda$.  However, it has a serious drawback: one finds an explicit 
RG scale dependence in the dimensionless RG flow equations and it makes no room for non-trivial 
fixed points (only pseudo fixed points can be identified).

In Ref.~\cite{thermal_rg_cosmo} we proposed to reconcile points i) and ii) by relating the 
temperature parameter of the finite-temperature formalism to the running RG scale
\beq
\label{relation}
T \equiv k_T = \tau k
\eeq
where $k_T$ and $k$ serve as the running cutoff for thermal and quantum fluctuations, respectively. 
In this approach, the dimensionless variable factor $\tau$ is kept constant over the RG flow and it 
plays the role of the temperature of the cosmic plasma. By this choice we showed that the 
dimensionless RG flow equations have no explicit RG scale-dependence \cite{thermal_rg_cosmo}. 
Thus, one can find non-trivial fixed points and the usual RG flow diagram method can be used to 
study critical behaviour \cite{thermal_rg_qpt_cpt}. We also showed how this modified thermal RG 
equation was used to "solve the triviality" of the $\phi^4$ scalar field theory, i.e., the Higgs-like 
inflationary model.
The "triviality" means that in $d=4$ dimensions the $\phi^4$ model has two phases but the classical 
and quantum analysis gives the same separatrix which is a vertical line at the vanishing mass.
The thermal RG flow equations suggested by us modifies the flow diagram and with non-vanishing 
value for $\tau$, the RG trajectory which separates the phases is no longer a vertical line which
"solves the triviality" of the model.
However, the infinite temperature limit cannot be taken unambiguously without taking into 
account quantum gravity effects which become relevant at the Planck scale. This analysis 
was missing in \cite{thermal_rg_cosmo}. 

In this work, our goal is to understand how does the inclusion of gravity affects the proposed 
scheme. Thus, we apply the thermal FRG approach with the relation \eq{relation} on the AS 
quantum gravity with the Einstein-Hilbert truncation and determine its thermal RG flow in terms 
of $\tau$ which is considered as the temperature of the cosmic plasma.

\section{AS quantum gravity at zero temperature}
As a first step let us summarise the cornerstones of AS quantum gravity at zero temperature
starting from its simplest realization which is the Einstein-Hilbert truncation of the effective 
average action
\begin{equation}
\label{EH}
\Gamma_k = \frac{1}{16\pi G_k} \int d^4 x \, \sqrt{g} \, 
(2\Lambda_k-R) \,,
\end{equation}
where $g$ is the determinant of the metric tensor, $R$ is the Ricci scalar
and the scale-dependent parameters are the cosmological constant 
$\Lambda_k$ and the Newton coupling $G_k$.

\begin{figure}[t!] 
\begin{center} 
\epsfig{file=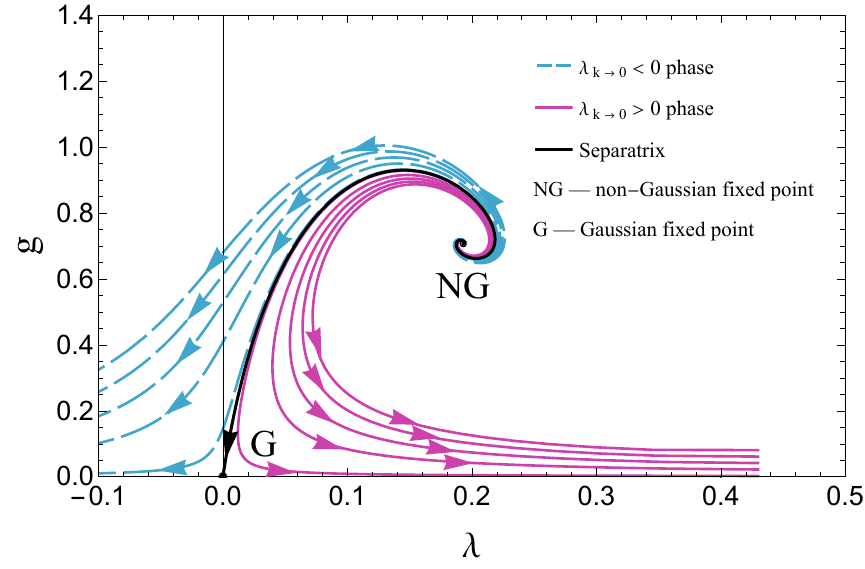,width=8.0cm}
\caption{\label{fig1} 
Zero-temperature RG flow diagram of AS quantum gravity obtained by the FRG 
equations with the Litim regulator using the Einstein-Hilbert truncation \eq{EH}.
} 
\end{center}
\end{figure}

The FRG study of \eq{EH} is given in terms of dimensionless couplings, 
$\lambda_k \equiv \Lambda_k k^{-2}$, $g_k \equiv G_k k^2$ with the help of the 
$\beta$-functions, see for example \cite{cosmo_4}
\beq
\nonumber 
k \partial_k \lambda_k = \beta_{\lambda}(\lambda_k, g_k), \qquad
k \partial_k g_k = \beta_g(\lambda_k, g_k) 
\eeq
which are calculated by the Litim regulator \cite{Litim2000}
\begin{align}
&\beta_g = (2 + \eta_N) g_k, \nn
&\beta_\lambda = (\eta_N -2) \lambda_k 
+ \frac{g_k}{12\pi} \left[ \frac{30}{1-2\lambda_k} 
- 24 - \frac{5}{1-2\lambda_k} \eta_N \right], \nonumber
\end{align}
where the anomalous dimension of the operator $\int d^dx \sqrt{g}R$, 
i.e., the Newton constant, $\eta_N = G_k^{-1} \, k\partial_k G_k$ 
is given by
\begin{equation}
\eta_N = \frac{g_k \, B_1}{1 - g_k \, B_2} \,,
\end{equation}
where
\bea
B_1 = \frac{1}{3\pi}\left[ \frac{5}{1-2\lambda_k} -
\frac{9}{(1-2\lambda_k)^2}  - 7 \right], \nn 
B_2 = 
- \frac{1}{12\pi}\left[ \frac{5}{1-2\lambda_k} -
\frac{6}{(1-2\lambda_k)^2} \right].
\eea
Based on these $\beta$-functions, the RG flow diagram is plotted on \fig{fig1}.
%
%
The model has two phases separated by the black RG trajectory which starts from the
non-Gaussian (NG) to the Gaussian (G) fixed point. The asymptotic safety is guaranteed
by the NG also called the Reuter fixed point \cite{reuter}. The separatrix can be 
interpreted similarly to that of the phase-transition line in a $\phi^4$ theory. In that case one can 
distinguish two different phases based on the sign of the mass term; a symmetric phase with 
$m^2 >0$ and a symmetry broken one where $m^2<0$ applies. In the Einstein-Hilbert action 
$\Lambda$ is positive, however the RG equations are well defined in the case of $S \to -S$ as 
well \cite{reuter}, this is used to trade the "wrong sign" of the kinetic term for an upside down 
potential \cite{reuter_weyer_2009_1}. This transformation results in a negative $\Lambda_k$ 
term in the scale dependent effective action, therefore the phase which results in a negative 
cosmological contant in the IR limit, i.e., $\lambda_{k\to 0} <0$ can be seen as the symmetric 
one, while the phase with $\lambda_{k\to 0} >0$ can be interpreted as the symmetry broken. 
In this work we apply this terminology which was also used in Ref. \cite{flora} to differentiate 
between the two phases, where the cosmological constant takes on a different sign. 
Thus, the symmetric (i.e., $\lambda_{k\to 0} <0$) phase is indicated by dashed blue RG 
trajectories while the broken (i.e., $\lambda_{k\to 0} >0$) phase is given by pink solid lines. 
A similar situation can be seen for conformally reduced AS gravity
\cite{reuter_weyer_2009_1,CREH1,CREH2,CREH3,reuter_weyer_2009_2,bonanno_2023}.

\begin{figure}[t!] 
\begin{center} 
\epsfig{file=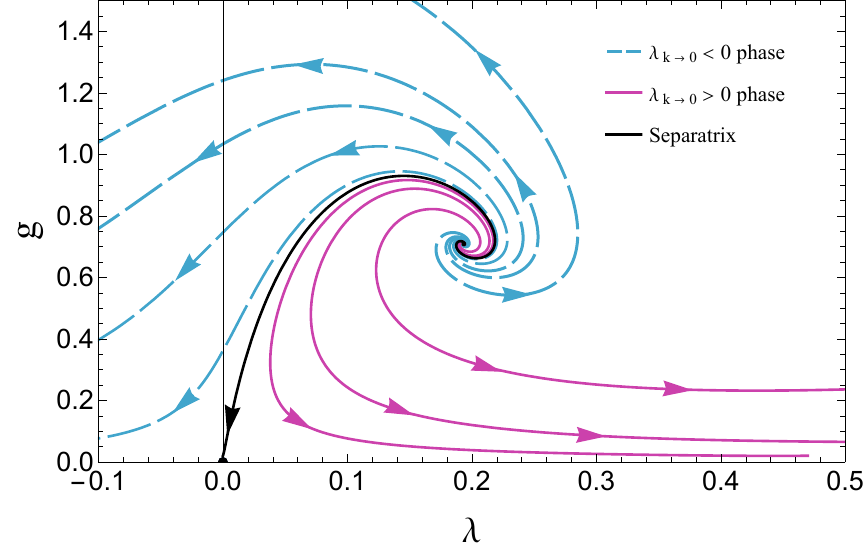,width=8.0cm}
\caption{\label{fig2} 
Identical to \fig{fig1} but obtained by the numerical solution of the general $\Fbeta$-functions 
\eq{general_beta_1} and \eq{general_beta_2} with the Litim regulator. 
} 
\end{center}
\end{figure}

In order to study the model at finite temperatures, the $\Fbeta$-functions \cite{AS_grav_1},
\beq
\nonumber 
k \partial_k \lambda_k = \Fbeta_{\lambda}(\lambda_k, g_k), \qquad
k \partial_k g_k = \Fbeta_g(\lambda_k, g_k) 
\eeq
must be given by their most general form  including the so 
called threshold functions \cite{AS_grav_1},
\begin{align}
\label{general_beta_1}
&\Fbeta_g(\lambda_k, g_k) = \left(d-2+\eta_N \right) \, g_k  \\
\label{general_beta_2}
&\Fbeta_{\lambda}(\lambda_k, g_k) = -(2-\eta_N)\, \lambda_k + \frac{1}{2}\, (4 \pi)^{1-d/2}  \, g_k \\
&\big[ 2 \, d(d+1) \, \Phi^1_{d/2}(-2\lambda_k)- 8 \, d \, \Phi^1_{d/2}(0)  \nonumber \\
&- d(d+1) \,
\eta_N \, \widetilde{\Phi}^1_{d/2}(-2 \lambda_k) \big] \nonumber
\end{align}
where the anomalous dimension $\eta_N$ is given by \cite{AS_grav_1}
\beq
\eta_N(g_k, \lambda_k) = \frac{g_k \, B_1(\lambda_k)}{1-g_k \, B_2(\lambda_k)},
\eeq
and the functions $B_1(\lambda_k)$ and $B_2(\lambda_k)$ have the following definition
\begin{align}
&B_1(\lambda_k) = \frac{1}{3} \,(4 \pi)^{1-d/2} \bigg[ d(d+1) \, \Phi^1_{d/2-1}(-2\lambda_k) \\
&- 6d(d-1)\,\Phi^2_{d/2}(-2\lambda_k) - 4d \,\Phi^1_{d/2-1}(0) - 24 \Phi^2_{d/2}(0) \bigg] \nn
&B_2(\lambda_k) = -\frac{1}{6} \,(4 \pi)^{1-d/2} \bigg[d(d+1) \,\widetilde{\Phi}^1_{d/2-1}(-2\lambda_k) \nn
&-6d(d-1)\widetilde{\Phi}^2_{d/2}(-2\lambda_k)  \bigg], 
\end{align}
where we introduced the (dimensionless) regulator $R^{(0)}$, i.e., the shape function 
which appears in threshold functions $\Phi^{p}_{n}$ and $\widetilde{\Phi}^{p}_{n}$ as \cite{AS_grav_1}
\begin{align}
\label{threshold_func_1}
\Phi^p_n(w) &=& \frac{1}{\Gamma(n)} \int^{\infty}_{0} \!\! dz z^{n-1} \frac{ R^{(0)}(z) \, - \, z\, R^{(0)\prime}(z)}{\left[ z +  R^{(0)}(z) + w\right]^p} \\
\label{threshold_func_2}
\widetilde{\Phi}^p_n(w) &=& \frac{1}{\Gamma(n)} \int^{\infty}_{0} \!\! dz z^{n-1} \frac{ R^{(0)}(z) }{\left[ z +  R^{(0)}(z) + w\right]^p},
\end{align}
with $p=1,2$. For a simple consistency check, on \fig{fig2} we plotted the RG flow diagram 
by using these general $\Fbeta$-functions with the Litim regulator \cite{Litim2000} and
we obtained results identical to \fig{fig1}.
%
%
%

\section{Thermal RG study of AS quantum gravity}
In order to extend the zero-temperature RG study of the AS quantum gravity for finite temperatures 
one has to modify the threshold functions \eq{threshold_func_1} and \eq{threshold_func_2} but 
nothing else. This modification is required because the upper bound of the imaginary time $\tilde t$ 
integral $\int d^dx \to \int_0^\beta d\tilde t \int d^{d-1}x$ becomes finite where $\beta = 1/T$ and $T$ 
is the temperature parameter. As a result, the momentum integral with respect to the imaginary time is 
replaced by a summation
\beq
\label{matsubara}
\int d^dp \to T \sum_{\omega_m} \int d^{d-1}p
\eeq
over the Matsubara frequencies; for bosonic degrees of freedom $\omega_m = 2 \pi m T$,
where $m \in \mathbb{Z}$. In order to apply \eq{matsubara} to the zero-temperature threshold functions 
\eq{threshold_func_1} and \eq{threshold_func_2}, one should transform them back to momentum integrals, 
implement the Matsubara summation on one of them, and then perform again the solid angle integrals. 
This implies the following changes to \eq{threshold_func_1} and \eq{threshold_func_2} 
\begin{align}
\label{threshold_func_1_T}
\Phi^p_n(w,\tau) = \frac{ \, 2\tau \sqrt{\pi} }{\Gamma\left( n-\frac{1}{2} \right)}\ \sum_{m \ = \  -\infty}^{\infty} \int^{\infty}_{0} dy \ y^{n-\frac{3}{2} }  \nn
\frac{R^{(0)}(y) - y \,  R^{(0)\prime}(y)}{\left[ y + (2 m \pi \tau)^2 + R^{(0)}(y) + w \right]^p}, \\
\label{threshold_func_2_T}
\widetilde{\Phi}^p_n(w,\tau) = \frac{ \, 2\tau \sqrt{\pi} }{\Gamma\left( n-\frac{1}{2} \right)}\ \sum_{m \ = \  -\infty}^{\infty} \int^{\infty}_{0} dy \ y^{n-\frac{3}{2} }  \nn
\frac{R^{(0)}(y)}{\left[ y + (2 m \pi \tau)^2 + R^{(0)}(y) +w \right]^p},
\end{align}
where we used the notation $z \equiv y$ and introduced the dimensionless temperature 
\eq{relation}, i.e., $\tau = T/k \equiv k_T/k$. It is important to note that $\tau$ is kept constant 
over the RG flow.

\begin{figure}[t!] 
\begin{center} 
\epsfig{file=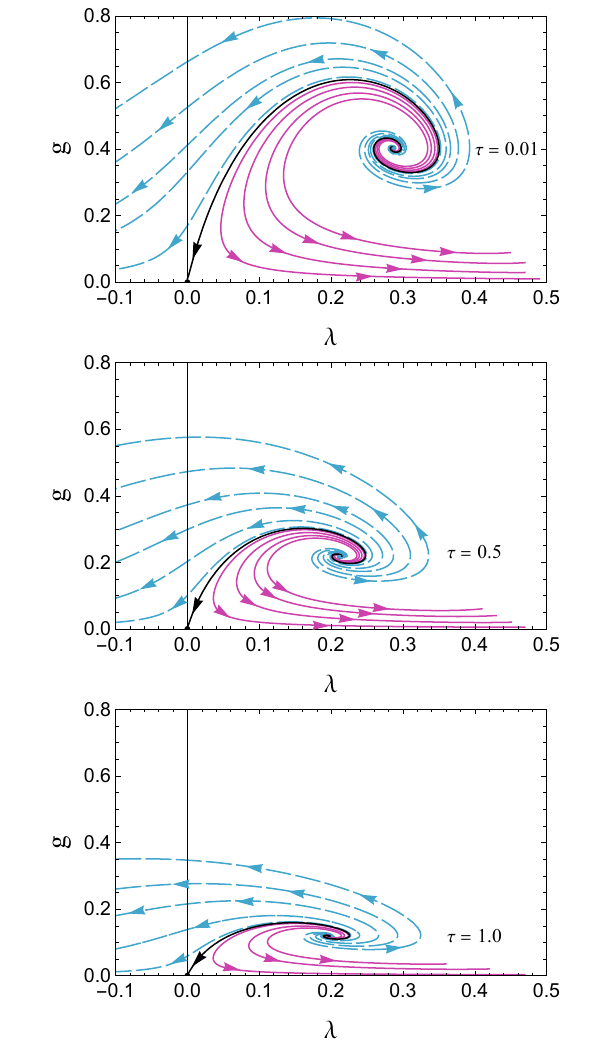,width=8.0cm}
\caption{\label{fig3} 
Thermal RG flow diagrams of AS quantum gravity for various values of the dimensionless 
temperature $\tau$. In the high temperature limit, i.e., $\tau \to \infty$, the $g$-coordinate of 
the Reuter fixed point tends to zero.
} 
\end{center}
\end{figure}

\begin{figure}[t!] 
\begin{center} 
\epsfig{file=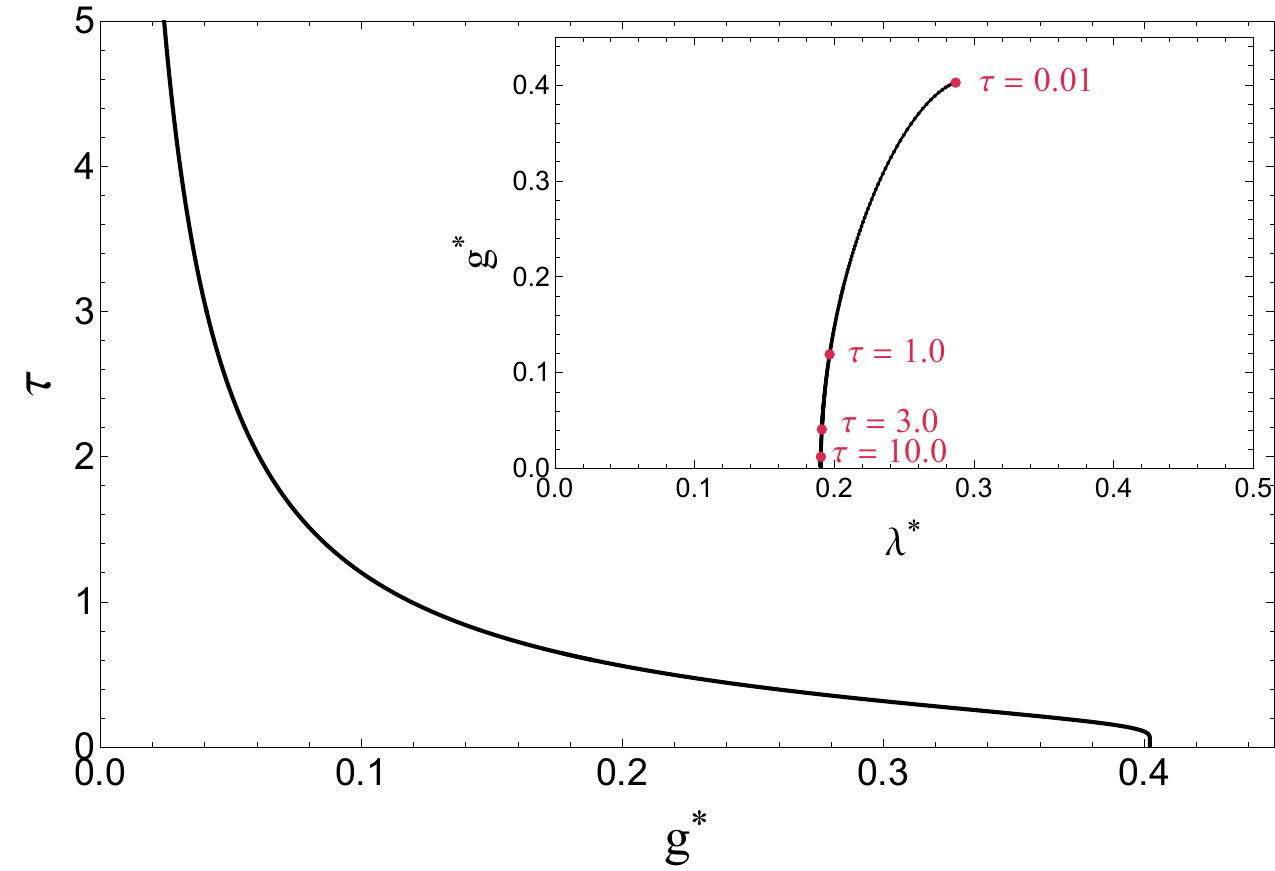,width=8.0cm}
\caption{\label{fig4} 
QPT-CPT diagram of AS quantum gravity in terms of the dimensionless temperature $\tau$
and the $g$-coordinate of the Reuter fixed point. The black line, i.e., the function $g^\star(\tau_c)$, 
is the critical line which separates the symmetric and the broken phases of the model. For a 
given (but fixed) $g^\star$-value, for $\tau > \tau_c$ or $\tau < \tau_c$ model is in the 
symmetric or its broken phase. The inset shows how the positions of the Reuter fixed point 
($g^\star$, $\lambda^\star$) changes by $\tau$. 
} 
\end{center}
\end{figure}

Thermal RG flow diagrams of AS quantum gravity for various values of the dimensionless 
temperature $\tau$ are plotted on \fig{fig3}. In the zero-temperature case at 
$\lambda = 1/2$ singularities are present in the beta functions, causing the trajectories to 
terminate. This pole persists for finite temperatures as well, one can see on \fig{fig3} the 
trajectories stop at a certain point in the $\lambda>0$ phase.
%
%
The subfigures suggest, that in the limit of high temperature, i.e., for $\tau \to \infty$, the 
$g$-coordinate of the Reuter fixed point, which we denote by $g^\star$, vanishes. This is 
confirmed by the inset of \fig{fig4} where the positions of the Reuter fixed point is plotted for 
various values of $\tau$.
%
%

Before we analyse the consequences of the large temperature behaviour let us first comment
the zero-temperature limit. It is clear from \fig{fig3} and \fig{fig4} that the position of the Reuter
fixed point in the limit $\tau \to 0$ {\em does not} coincide to its zero-temperature value, see \fig{fig1} 
or \fig{fig2} (keep in mind that \fig{fig1} and \fig{fig2} are identical). This is a consequence of the 
use frequency-dependent or independent regulator function in the calculation of threshold functions,
see for example the discussion in \cite{thermal_rg_qpt_cpt}. The finite-temperature formalism 
requires a frequency independent, i.e., cylindrically symmetric \cite{cylindrical_symm_1,
cylindrical_symm_2} regulator which does not regulate the Matsubara sum. This is not true for 
the zero-temperature case where momentum integrals (including the imaginary time direction) are 
performed by the regulator function i.e., with a frequency-dependent, i.e., spherically symmetric 
regulator. This is the reason why, the zero-temperature limit of truncated thermal FRG equations 
may differ from their counterparts at zero temperature.

For intermediate temperatures, i.e., $0 < \tau < \infty$, one can observe a thermal or classical 
phase transition (CPT) in the following way. Let us choose, arbitrary (but positive) initial values for 
the couplings in the vicinity of the Gaussian fixed point as a starting point. Then, one can always 
define a particular (dimensionless) temperature, $\tau_c$, which results in a separatrix, see 
e.g., black lines of subfigures in \fig{fig3}, which goes through this particular initial point. For 
large temperatures, $\tau > \tau_c$, the RG trajectory starting form this chosen initial point 
always ends up in the symmetric phase. For small temperatures, $\tau < \tau_c$, the RG 
trajectory from the initial point runs into the broken phase. Thus, the model undergoes a thermal
phase transition at $\tau = \tau_c$. 

\begin{figure}[t!] 
\begin{center} 
\epsfig{file=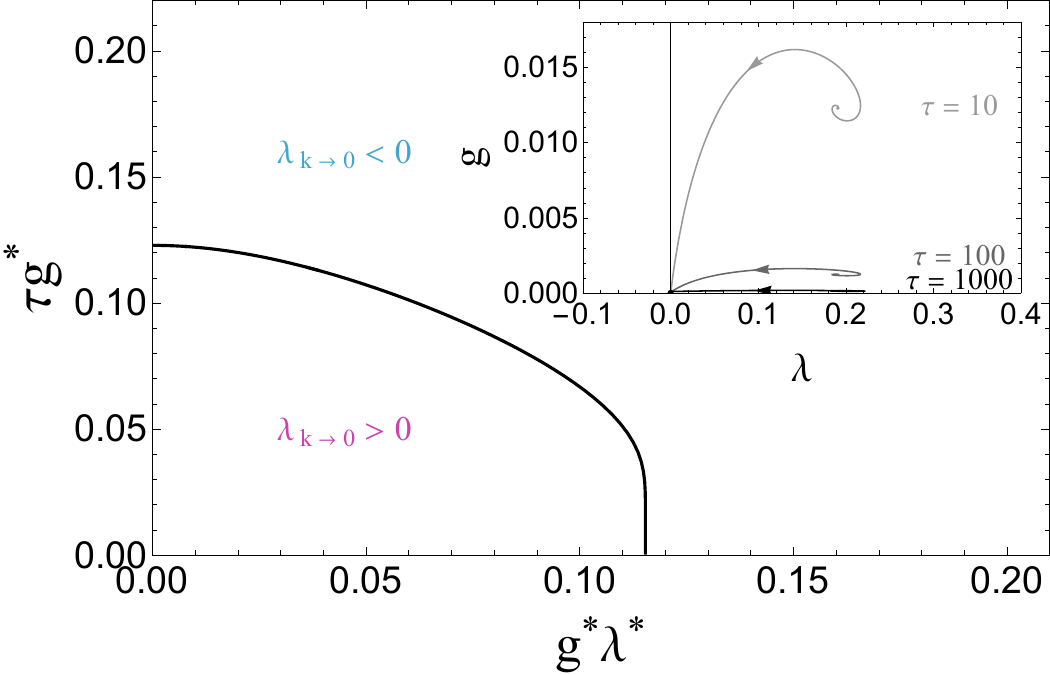,width=8.0cm}
\caption{\label{fig5} 
QPT-CPT diagram in terms of $\tau g^\star$ and $g^\star\lambda^\star$
where $g^\star$ and $\lambda^\star$ are the coordinates of the Reuter fixed point. 
The black critical line which separates the phases terminates at finite value in the limit of 
$g^\star\lambda^\star \to 0$ which represents an important difference compared to \fig{fig4} 
where the critical line runs to infinity. The inset shows how the separatrix looks like for large 
values of $\tau$.
} 
\end{center}
\end{figure}

For a fixed (dimensionless) temperature $\tau$, one can observe a quantum phase transition 
(QPT) depending on the choice of the initial values of the couplings, i.e., starting point chosen 
in the vicinity of the Gaussian fixed point. If this starting point is chosen to be above the separatrix
(the black trajectory connecting the G and NG fixed points), the RG trajectory from 
this starting point ends up in the symmetric phase, and if the starting point is below the separatrix, 
its RG trajectory runs to the broken phase of the model. For different separatrices at 
various large $\tau$ see the inset on \fig{fig5}.
Although, the slope of the separatrix serves as the real "quantum parameter" which controls the 
QPT, a good approximation for this could be the $g$-coordinate of the Reuter fixed point which
we denote by $g^\star$.

One can construct a function, $g^\star(\tau_c)$ which represents a critical line in the $\tau-g$
plane and separates the symmetric (above this line) and broken (below this line) phases of the
model. This critical line is plotted on \fig{fig4} which represents the so called QPT-CPT diagram
of the AS quantum gravity. It is illustrative to compare it to the QPT-CPT diagram of the $\phi^4$
model in lower dimensions, see Fig.~(5) of \cite{thermal_rg_qpt_cpt}. One finds many similarities.
For example, for $\tau \to 0$ critical lines always end up in the quantum critical point of the model.
Let us note that the QPT of the $\phi^4$ scalar field theory has also been investigated in 
connection to the Naturalness/Hierarchy problem \cite{BrBrCoDa,BrBrCo,qpt_higgs1,qpt_higgs2}. 
For example, in \cite{qpt_higgs1,qpt_higgs2} its was shown that the hierarchy problem as well as 
the metastability of the electroweak vacuum can be understood as the Higgs potential being 
near-critical, i.e., close to a QPT. Another similarity is that for vanishing "quantum parameter", 
the critical (dimensionless) temperature $\tau_c$ tends to infinity, however, as it is discussed in
\cite{thermal_rg_qpt_cpt} this limit is non-analytic: at zero "quantum parameter" the critical 
temperature must be zero. 

%
%
As discussed before, for $\tau \to \infty$ the $g^*$ disappears, i.e., $g^* \to 0 $, therefore we 
investigated their product ($\tau g^*$) and plotted it with respect to $g^*\lambda^*$, which 
serves as the "quantum parameter" of the model. The latter choice is motivated by the fact that 
$G_k \Lambda_k = g_k \lambda_k$ is a dimensionless combination of the couplings. It is worth 
noting, however, that the change in $\lambda^*$ during the thermal evolution is relatively small 
compared to that in $g^*$, at large $\tau$ one can view $\lambda^*$ as "constant". This implies, 
that e.g., the slope of the separatrix ($g^* /\lambda^*$) or $g^*$ quantum parameters produce 
qualitatively the same QPT-CPT diagram as of \fig{fig5}. We have found that as $g^* \lambda^*$ 
quantum parameter disappears, the $\tau g^*$ dimensionless parameter tends to 
a constant value. Thus, \fig{fig5} gives a more direct analogy to the known QPT-CPT diagrams 
of various statistical physical models (e.g., Ising model).

Finally let us come back to the discussion of the large temperature limit. As it is clearly shown
by the inset of \fig{fig4} (and also suggested by the subfigures of \fig{fig3}) for $\tau \to \infty$ 
the $\lambda$-coordinate of the Reuter fixed point goes to a positive value
and its $g$-coordinate vanishes, i.e., $g^\star(\tau_c\to \infty) = 0$. 
This means that for very large temperatures, larger than for example the Planck temperature
where quantum gravity effects must be taken into account, according to our framework, only
the symmetric phase of the AS quantum gravity survives which is the most important finding
of this work.

\section{Conclusions}

In this work, we applied a modified thermal FRG approach recently introduced in 
\cite{thermal_rg_cosmo} on the AS quantum gravity by using the Einstein Hilbert 
truncation in $d=4$ dimensions. The essence of this modified method was the use of a 
constant dimensionless temperature, $\tau = T/k$ where T was assumed to run with 
the RG scale $k$. We explored the possibility of this choice on the phase structure 
of AS quantum gravity. The focus was on the large temperature limit, i.e, $\tau \to\infty$,
in which quantum gravity effects must be taken into account.

Our most important finding was that for very large temperatures, only the symmetric phase of 
the AS quantum gravity survived because the $g$-coordinate of the Reuter fixed point vanished.
On the one hand it is a very natural expectation, that for very large temperatures the model 
is in its symmetric phase. On the other hand it is not in the usual/standard picture of quantum
gravity which is linked to behaviour of the model around UV fixed points. This is because at 
zero temperature, it is very natural to identify the running scale parameter $k$ with physical 
properties of the system and look for a particular RG trajectory picked up by Nature 
\cite{k_flow_1,k_flow_2,k_flow_3} which starts from the vicinity of the Gaussian fixed
point at very high energies and tends to low energies \cite{IR_flow_1,IR_flow_2,IR_flow_3}. 
However, at finite temperatures, it is more natural to associate physical properties of the model 
to its temperature ($\tau$) and the RG scale $k$ is considered only as a tool to take into account 
quantum effects which are fully incorporated in the limit $k \to 0$ only. In other words, 
we do not consider the k-flow as an approximation to catch the main physics properties of 
the system at a given range of energy.

One can notice that in case of large temperatures the cosmological constant takes on a negative 
value in the limit $k\to 0$. This was also initially predicted by certain string theories, which 
contradicts current cosmological observations prompting attempts to resolve the discrepancy 
\cite{kachru_2003}. However, these solutions might describe unstable worlds \cite{hertog_2003}. 
In our framework, it seems natural for high-energy (high-temperature) models to yield a negative 
cosmological constant. This is not in disagreement with observations, since during the thermal 
evolution of the Universe a phase transition occurs and the cosmological constant runs to the 
expected positive value at low temperatures.

Finally, let us draw the attention of the reader that all computations of the present work 
are performed in the Euclidean spacetime. There is an increasing interest in the literature 
to discuss whether an analytic continuation to Minkowski spacetime is possible in the framework of 
the Wilsonian RG method, see e.g., \cite{minkowski_frg, lorentz_1, lorentz_2, lorentz_3, lorentz_4}. 
However, we refer to lattice calculations performed in the Euclidean spacetime \cite{lattice}, as a 
strong support for the use of Euclidean instead of the Minkowski spacetime.

\section*{Acknowledgements} 
Supported by the University of Debrecen Program for Scientific Publication.

\end{document}